\documentclass[a4paper,11pt]{article}
\pdfoutput=1 

\usepackage[usenames,dvipsnames,svgnames,table]{xcolor} 
\usepackage{jcapmod}
\usepackage{verbatim}

\usepackage[T1]{fontenc} 
\usepackage{inputenc}
\definecolor{lightgray}{gray}{0.9}
\definecolor{Amber}{rgb}{1.0, 0.75, 0.0}
\definecolor{blizzardblue}{rgb}{0.67, 0.9, 0.93}

\usepackage{float}
\usepackage{graphicx}
\usepackage{hyperref}
\usepackage{amsmath}
\usepackage{amssymb}
\usepackage{microtype}
\usepackage[shortlabels]{enumitem}
\usepackage{cancel}
\usepackage{amsbsy}
\usepackage{color}
\usepackage[capitalise]{cleveref}
\usepackage{breakurl}
\usepackage{mathtools}
\usepackage{csquotes}
\usepackage{lmodern}
\usepackage{bm}
\usepackage{dsfont}
\usepackage{bbold}
\usepackage{empheq}

\usepackage{caption}
\usepackage{subcaption}
\usepackage{placeins}
\usepackage{xspace}

\pdfpageheight\paperheight
\pdfpagewidth\paperwidth

\usepackage[textwidth=0.71\paperwidth, textheight=0.8\paperheight]{geometry}
\usepackage[table]{xcolor}

\newcommand{\dderiv}{\mathrm{d}}
\newcommand{\EB}{\emph{EB}\xspace}
\newcommand{\Clebsch}{\mathcal{C}}
\newcommand{\vecnp}{\smash{\vec{n}'}}
\renewcommand*{\vec}[1]{\bm{#1}}
\newcommand*{\unitvec}[1]{\vec{\hat{#1}}}
\newcommand*{\mat}[1]{\bm{\mathsf{#1}}}
\newcommand*{\E}[1]{\texorpdfstring{\ensuremath{E_{#1}}}{E#1}}
\newcommand*{\Espace}{\texorpdfstring{\ensuremath{E^3}}{E(3)}}
\let\field\phi
\let\phi\varphi

\usepackage{dsfont}
\allowdisplaybreaks

\makeatletter
\DeclareRobustCommand{\rcite}[1]{%
  \rcite@aux#1,\@nil{#1}%
}
\def\rcite@aux#1,#2\@nil#3{%
  \if\relax#2\relax
    Ref.~\cite{#3}%
  \else
    Refs.~\cite{#3}%
  \fi
}
\makeatother

\subheader{IFT-UAM/CSIC-24-104}

\title{\boldmath Cosmic topology. Part IIIa. Microwave background parity violation without parity-violating microphysics}

\author[a]{Amirhossein Samandar,}
\author[b]{Javier Carr\'on Duque,}
\author[a]{Craig J. Copi,}
\author[b]{Mikel Martin Barandiaran,}
\author[a]{Deyan P. Mihaylov,}
\author[c]{Thiago S. Pereira,}
\author[a]{Glenn D. Starkman,}
\author[b,a,d]{Yashar Akrami,}
\author[e,f,g]{Stefano Anselmi,}
\author[a]{Fernando Cornet-Gomez,}
\author[h]{Johannes R. Eskilt,}
\author[d]{Andrew H. Jaffe,}
\author[i]{Arthur Kosowsky,}
\author[a]{and Andrius Tamosiunas}

\collaboration{(COMPACT Collaboration)}
\collaborationImg{\includegraphics[width = 0.1\textwidth]{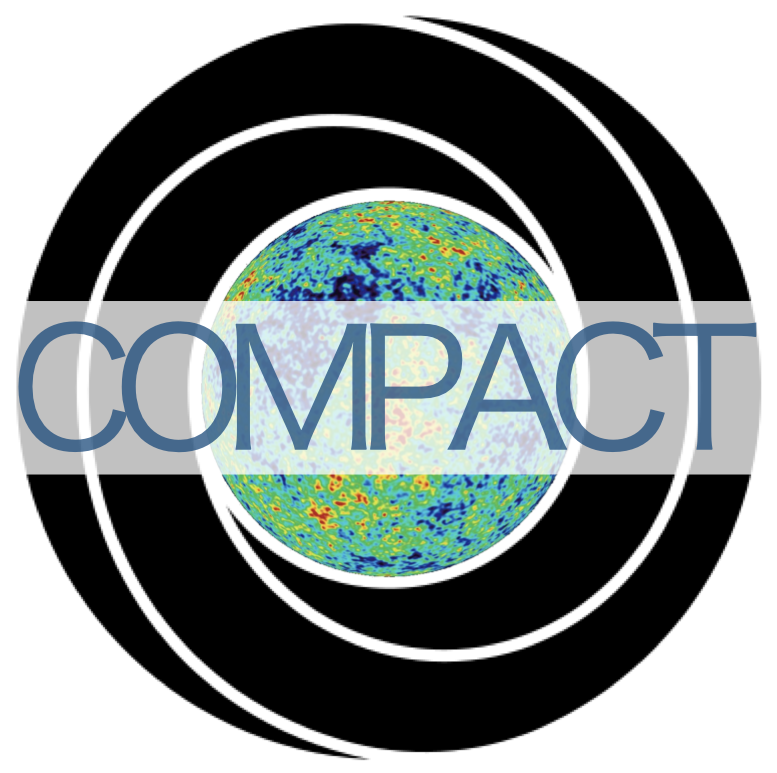}}

\affiliation[a]{CERCA/ISO, Department of Physics, Case Western Reserve University, 10900 Euclid Avenue, Cleveland, Ohio 44106, USA}
\affiliation[b]{Instituto de F\'isica Te\'orica (IFT) UAM-CSIC, C/ Nicol\'as Cabrera 13-15, Campus de Cantoblanco UAM, 28049 Madrid, Spain}
\affiliation[c]{Departamento de F\'{i}sica, Universidade Estadual de Londrina,
Rod. Celso Garcia Cid, Km 380, 86057-970, Londrina, Paran\'a, Brazil}
\affiliation[d]{Astrophysics Group \& Imperial Centre for Inference and Cosmology, Department of Physics, Imperial College London, Blackett Laboratory, Prince Consort Road, London SW7 2AZ, United Kingdom}
\affiliation[e]{INFN, Sezione di Padova, via Marzolo 8, I-35131 Padova, Italy}
\affiliation[f]{Dipartimento di Fisica e Astronomia ``G. Galilei'', Universit\`a degli Studi di Padova, via Marzolo 8, I-35131 Padova, Italy}
\affiliation[g]{Laboratoire Univers et Th\'eories, Observatoire de Paris, Universit\'e PSL, Universit\'e Paris Cit\'e, CNRS, F-92190 Meudon, France}
\affiliation[h]{Institute of Theoretical Astrophysics, University of Oslo, P.O. Box 1029 Blindern, N-0315 Oslo, Norway}
\affiliation[i]{Department of Physics and Astronomy, University of Pittsburgh, Pittsburgh, Pennsylvania 15260, USA}

\emailAdd{amirhossein.samandar@case.edu}
\emailAdd{javier.carron@csic.es}
\emailAdd{craig.copi@case.edu}
\emailAdd{mikel.martin@uam.es}
\emailAdd{deyan.mihaylov@case.edu}
\emailAdd{tspereira@uel.br}
\emailAdd{glenn.starkman@case.edu}
\emailAdd{yashar.akrami@csic.es}
\emailAdd{stefano.anselmi@pd.infn.it}
\emailAdd{fernando.cornetgomez@case.edu}
\emailAdd{j.r.eskilt@astro.uio.no}
\emailAdd{a.jaffe@imperial.ac.uk}
\emailAdd{kosowsky@pitt.edu}
\emailAdd{andrius.tamosiunas@case.edu}

\date{\today}

\abstract{The standard cosmological model, which assumes statistical isotropy and parity invariance, predicts the absence of correlations between even-parity and odd-parity observables of the cosmic microwave background (CMB). 
Contrary to these predictions, large-angle CMB temperature anomalies generically involve correlations between even-$\ell$ and odd-$\ell$ angular power spectrum $C_\ell$, while recent analyses of CMB polarization have revealed non-zero equal-$\ell$ \EB correlations. 
These findings challenge the conventional understanding, suggesting deviations from statistical isotropy, violations of parity, or both. 
Cosmic topology, which involves changing only the boundary conditions of space relative to standard cosmology, offers a compelling framework to potentially account for such parity-violating observations. 
Topology inherently breaks statistical isotropy, and can also break homogeneity and parity, providing a natural paradigm for explaining observations of parity-breaking observables without the need to add parity violation to the underlying microphysics. 
Our investigation delves into the harmonic space implications of topology for CMB correlations, using as an illustrative example \EB correlations generated by tensor perturbations under both parity-preserving and parity-violating scenarios. 
Consequently, these findings not only challenge the foundational assumptions of the standard cosmological model but also open new avenues for exploring the topological structure of the Universe through CMB observations.}

\keywords{cosmic topology, cosmic microwave background, polarization, parity violation, statistical isotropy, cosmic anomalies}

\begin{document}
\maketitle
\flushbottom

\section{Introduction}
\label{sec:introduction}

The observed Universe is neither homogeneous nor isotropic but appears to be nearly both \cite{Planck:2013lks,Planck:2015igc,Planck:2019evm}. 
Our contemporary theory of cosmology, therefore, describes the Universe in terms of a spatially maximally symmetric Friedmann-Lema\^{i}tre-Robertson-Walker (FLRW) spacetime, i.e., with a spatially homogeneous and isotropic background metric, filled with homogeneous and isotropic sources of stress-energy. 
On top of this background, there are small fluctuations in both the local geometry and stress-energy tensor. 
All the underlying microphysics, i.e., the Lagrangian, is taken to be invariant under arbitrary translations, rotations, and, at least in the gravitational/geometric sector, parity transformations (i.e., spatial inversions). 
Therefore, the statistical properties of the fluctuations also respect these symmetries: while any particular realization of these fluctuations breaks isotropy, parity, and homogeneity, they arise from an underlying statistical process that respects these symmetries.

In this description, spatial homogeneity and isotropy play a central role. 
When combined with physics which preserves parity at the level of the Lagrangian, these symmetries force most auto- and cross-correlations of observables to vanish in harmonic space. 
Conversely, and crucially for this paper,
the breaking of spatial symmetries leads to a proliferation of expected correlations that are forbidden in a globally homogeneous, isotropic Universe, even if the symmetries are preserved by the microphysics.
This is what happens, for example, when we consider cosmic topology, which breaks spatial symmetries at the level of the boundary conditions \cite{COMPACT:2022gbl, COMPACT:2022nsu, COMPACT:2023rkp}. 
This is broadly analogous to the phenomenon of spontaneous symmetry breaking.

Observations of the cosmic microwave background (CMB) provide valuable insights into these correlations. 
The CMB exhibits not only temperature anisotropies ($T$) but also polarization patterns characterized by scalar (\emph{E}-mode) and pseudo-scalar (\emph{B}-mode) fields on the sky. 
Using these fields, we can construct various parity-even and parity-odd correlations between pairs of spherical harmonic coefficients $a_{\ell m}^X$ and $a_{\ell' m'}^Y$, with $X, Y\in\{T, E, B\}$.
If the fluctuations are statistically isotropic and homogeneous, and preserve parity, then all \emph{primordial} parity-odd correlations vanish \cite{Weinberg:2008}---the detected correlations being attributed to secondary effects such as gravitational lensing~\cite{Hu:2000ee,Schneider:2001af}.

However, the assumption of statistical isotropy has long been questioned based on the observed large-scale anomalies discovered in the CMB temperature data~\cite{Planck:2013lks,Schwarz:2015cma,Planck:2015igc,Planck:2019evm,Abdalla:2022yfr}. Recently, several of us, using four representative statistics from different classes of these anomalous statistics, have characterized the correlations among these anomalies, discovering that collectively they provide $>\!5\sigma$ evidence for the violation of statistical isotropy on large scales~\cite{Jones:2023ncn}.
This is overwhelming evidence that the CMB is \emph{not} the result of the statistically isotropic (very nearly) Gaussian process envisaged in current models. Additionally, observations of the galaxy distribution also seem to be at tension with the assumption of homogeneity and isotropy, as in the dipole of the galaxy distribution (apparently incompatible with the kinematic dipole extracted from the CMB at the $4.9\sigma$ confidence level \cite{Secrest:2020has}, though this is still controversial \cite{Maartens:2017qoa,Dalang:2021ruy,Guandalin:2022tyl,vonHausegger:2024jan}) or the large-scale bulk flow~\cite{Watkins:2023rll}. 

Parity-violating \EB correlations may also have been detected in the CMB. It has long been known (see, e.g., \rcite{Hu:2000ee,Lewis:2006fu}) that CMB lensing can convert \emph{E}-mode polarization to \emph{B}-mode polarization, generating a connected four-point function that, according to the {\it Planck} 2018 data release \cite{Planck:2018vyg}, can now be measured. Though not directly related to a cosmological \EB two-point correlation, it led to a hint of parity violation based on a simplified treatment of polarized Galactic dust emission where a 99.2\% ($2.4\sigma$) confidence level preference for a non-zero \EB signal was found \cite{Minami:2020odp}. 
After a more careful accounting of the polarized dust emission from the Milky Way, the reported significance of the potential \EB correlation increased to 99.987\% ($3.6\sigma$) confidence level and the signal  was found to be frequency-independent \cite{Komatsu:2022nvu,Eskilt:2022wav,Eskilt:2022cff}. 
While not yet at the conventional $5\sigma$ threshold for a claimed discovery, this emphasized the importance of and potential for measuring \EB correlations. 

Several mechanisms to produce \EB correlations breaking parity at the level of the microphysics have been proposed in the literature. One possibility is cosmic birefringence, where the Universe contains a parity-violating field, typically either dark matter or dark energy.
This field interacts differently with the right- and left-handed states of photons, which causes the plane of polarization of photons to rotate.
This, in turn, produces a non-zero \EB correlation \cite{Minami:2019ruj,Diego-Palazuelos:2022dsq}. Another proposal is to break parity in the vacuum fluctuation during inflation to produce primordial chiral gravitational waves (CGWs).
These would lead to an imbalance between right- and left-handed circular polarization modes of the CGWs, effectively breaking parity symmetry and generating non-zero \EB correlations \cite{Cai:2016ihp,Takahashi:2009wc,Fujita:2022qlk}.

While parity-violating microphysics remains a viable mechanism for generating \EB correlations, the existence of non-zero \EB correlations does not necessarily imply parity-violating microphysics---when the off-diagonal sum of angular momentum modes satisfies $\ell + \ell'$ even, correlations between \emph{E}- and \emph{B}-modes are forbidden only by statistical isotropy. 
A violation of statistical isotropy can therefore lead to non-zero parity-conserving \EB correlations~\cite{Pitrou:2015iya} whenever the microphysics can generate both \emph{E} and \emph{B}.
For example, several works have explored the effects of anisotropic inflation models on CMB cross-correlations, as these models can introduce statistical anisotropy in the early Universe \cite{Pitrou:2008gk,Jazayeri:2014nya,Akhshik:2014gja}.

Non-trivial cosmic topology provides a plausible mechanism for breaking statistical isotropy, homogeneity, and in some cases, also parity. 
In this work, we show how, by changing the boundary conditions on tensor (spin-2) modes in this symmetry-breaking way, cosmic topology induces mixing between the \emph{E}-modes and \emph{B}-modes that would otherwise not be possible in a simply-connected, isotropic universe. In other words, cosmic topology generates \emph{EB} correlations in tensor-induced polarization. It similarly induces \emph{TB} correlations. In future work we will address \emph{TE} and \emph{TB} correlations from tensor modes \cite{COMPACT3b} and \emph{TE} correlations from scalar modes \cite{COMPACT3c} in a single comprehensive framework. The focus of this paper is on the impact of symmetry breaking on CMB correlations, with \emph{EB} correlations serving as a key signature of these broken symmetries in a non-trivial cosmic topology.

We demonstrate explicitly how the breaking of each symmetry leads to specific patterns of \EB correlation. 
The specific form of the \EB correlations depends on the topological shape, but their detection would provide a powerful signature of cosmic topology distinct from the imprints on temperature anisotropies alone.

This paper is organized as follows. In \cref{sec:symmetries} we provide a detailed overview of the mathematical implications of isotropy and parity invariance in the auto- and cross-correlations of observables in harmonic space, emphasizing which correlations must necessarily vanish and which in principle need not. In \cref{sec:topology} we review the interplay between non-trivial topologies and global homogeneity, isotropy, and parity invariance, developing some calculations in the 3-torus to showcase the emergence of correlations that are forbidden in the standard $\Lambda$CDM paradigm. Finally, in \cref{sec:results} we show the full tensor-fluctuation-induced \emph{EE}, \emph{BB}, and \EB correlation matrices for instances of the first three Euclidean topologies (\E{1}, \E{2}, and \E{3}), demonstrating that the gradual loss of symmetries translates into a far richer correlation structure; we also show that these correlations can induce a distinguishable signal for an ideal experiment.

\section{Symmetries and correlations}\label{sec:symmetries}

Before discussing how topology leads to \EB correlations, we first present a novel summary of the role of symmetries in restricting possible correlations (see \rcite{Mitsou:2019ocs,Abramo:2010gk,Lacasa:2013rbw} for related work). To understand the role of symmetries, and parity in particular, consider random fields on the sphere $\field^{X}(\unitvec{\Omega})$, requiring that they are 
either scalars or pseudo-scalars under parity.\footnote{
    For application to the CMB, we only require pure scalar or pseudo-scalar fields under parity.
    A general rotational scalar field can always be decomposed into a sum of a scalar field and a pseudo-scalar field under parity, and the argument presented here is carried through in a similar manner.}
Here $\unitvec{\Omega}$ is the location on the sphere and $X$ labels the field.

\subsection{The role of isotropy}

We first require these random fields to be statistically invariant under rotations, meaning their probability distribution function does not change under arbitrary rotations. Formally, under an $SO(3)$ transformation (i.e., a rotation) that takes $\unitvec{\Omega} \equiv (\theta,\phi)$ to $\unitvec{\Omega}' \equiv (\theta', \phi')$, the rotated field is $\field'^{X}(\unitvec{\Omega}) \equiv \field^{X}(\unitvec{\Omega}')$. Thus,
\begin{equation}
    \field'^{X}
    \stackrel{d}{=} \field^{X}
    \,,
\end{equation}
where $\stackrel{d}{=}$ denotes fields with the same probability distribution. Therefore, the statistical quantities derived from it, like the expected value or the variance, are also invariant under these transformations. This is physically motivated by the assumption that the Universe is statistically isotropic and homogeneous (i.e., statistically invariant under rotations and translations).

It is convenient to expand the random fields in spherical harmonics as
\begin{equation}
    \field^{X}(\unitvec{\Omega}) = \sum_{\ell = 0}^{\infty} \sum_{m=-\ell}^\ell \field^{X}_{\ell m} Y_{\ell m}(\unitvec{\Omega}) \,.
\end{equation}
The expectation value (i.e.,  ensemble average) of the coefficients with $\ell>0$ must vanish to respect the rotational symmetry of the background, since $\langle\field^{X}\rangle$ is invariant under rotations but only $Y_{00}$ possesses such a property.\footnote{More technically, each set $\{Y_{\ell m} \mid m=-\ell,...,\ell\}$ forms a $(2\ell+1)$-dimensional irreducible representation of $SO(3)$, and any quantity invariant under said group must therefore transform under the trivial, one-dimensional representation: $2\ell+1=1\implies \ell=0$. }
Thus, 
\begin{equation}
    \langle \field^{X}_{\ell m}  \rangle  \propto \delta_{\ell \,0}\, \delta_{m\,0} \,.
\end{equation}

Similarly, consider the expectation of the bilinear correlation of two fields measured in two different directions: $\langle \field^{X}(\unitvec{\Omega}_1) \field^{Y*}(\unitvec{\Omega}_2) \rangle$. In harmonic space, this reduces to expectation values of pairwise products of the field coefficients. Almost all such expectation values vanish if they respect the rotational symmetry of the background:
\begin{equation}
    \label{eqn:SICdiagonal}
    C^{XY}_{\ell m \ell' m'} \equiv 
	\langle \field^{X}_{\ell m} \field^{Y*}_{\ell' m'} \rangle =  C^{XY}_\ell \delta_{\ell \ell'}  \delta_{m m'} \,.
\end{equation}
This result arises because the product of two fields that are independently statistically isotropic is again statistically invariant under (double) rotations on $S^2 \times S^2$. 
We can represent the product $Y_{\ell m}(\unitvec{\Omega}_1) Y_{\ell' m'}(\unitvec{\Omega}_2)$ in a basis of total angular momentum using the bipolar spherical harmonics (BipoSH)~\cite{varshalovich1988quantum},
\begin{equation}
    \label{eqn:BipoSH}
   \{Y_\ell \otimes Y_{\ell'}\}_{LM}(\unitvec{\Omega}_1,\unitvec{\Omega}_2) = 
    \sum_{m=-\ell}^{\ell} \sum_{m'=-\ell'}^{\ell'} \Clebsch^{LM}_{\ell m \ell' m'} Y_{\ell m}(\unitvec{\Omega}_1)Y_{\ell' m'}(\unitvec{\Omega}_2),
\end{equation}
with $\Clebsch^{LM}_{\ell m \ell' m'}$ the Clebsch-Gordan coefficients. 
The $\{\{Y_\ell \otimes Y_{\ell'}\}_{LM} \mid M=-L, \ldots, L\}$ are the basis of a $(2L+1)$-dimensional irreducible representation of the rotation group $SO(3)$. Under a rotation $\mat{R}$, which carries $\unitvec{\Omega}_1 \to \unitvec{\Omega}'_1 \equiv \mat{R}\unitvec{\Omega}_1$ and $\unitvec{\Omega}_2 \to \unitvec{\Omega}'_2 \equiv \mat{R}\unitvec{\Omega}_2$, only the $(L,M)=(0,0)$ BipoSH 
is invariant under rotations.
Note that $L=0$ only appears in the sum on the right-hand side of \cref{eqn:BipoSH} if $\ell=\ell'$, and that $M=0$ necessarily implies $m'=-m$.\footnote{The $\delta_{m m'}$ in \cref{eqn:SICdiagonal} emerges rather than the $\delta_{-m m'}$ one might now expect because $\{Y_\ell \otimes Y_{\ell'}\}_{LM}$ is expressed in \cref{eqn:BipoSH} as a sum of product of two spherical harmonics, whereas $\langle \field^{X} \field^{Y*} \rangle$ involves products of a spherical harmonic and the complex conjugate of a spherical harmonic.}

In summary, statistical isotropy forces almost all cross-correlations between harmonic coefficients of observables to vanish to leading order in the small amplitudes of primordial fluctuations. It only allows correlations that are diagonal in harmonic space.
Conversely, if statistical isotropy is violated, values of the fields and their products need not be statistically invariant,
and $C^{XY}_{\ell m \ell' m'}$ from \cref{eqn:SICdiagonal} need not be diagonal.

\subsection{The role of parity}

It is a very common misconception (see, e.g.,~\rcite{baumann2022cosmology, Fabre:2013wia})  that invariance under parity transformations alone (which takes $\unitvec{\Omega}=(\theta,\phi)$ to $-\unitvec{\Omega}=(\pi-\theta,\pi+\phi)$) is sufficient to forbid all correlations between harmonic coefficients of parity-even observables (such as CMB temperature and \emph{E}-mode polarization) and parity-odd observables (such as \emph{B}-mode polarization).
This is not the case, as we will demonstrate.

Let $\field^{X+}$ and $\field^{Y+}$ represent random fields with even parity and $\field^{X-}$ and $\field^{Y-}$ represent random fields with odd parity (i.e., scalars and pseudo-scalars under parity, respectively).
Since the parity of the spherical harmonics is $(-1)^\ell$, the harmonic coefficients of these fields transform as
\begin{align}
    \field^{X+}_{\ell m} & \to (-1)^\ell \field^{X+}_{\ell m}, \\
    \field^{X-}_{\ell m} & \to (-1)^{\ell+1} \field^{X-}_{\ell m} . \nonumber
\end{align}
Thus, when two fields have the same parity, the bilinear correlations of these coefficients transform as
\begin{equation}
    \langle \field^{X\pm}_{\ell m} \field^{Y\pm *}_{\ell' m'} \rangle \to (-1)^{\ell+\ell'} \langle \field^{X\pm}_{\ell m} \field^{Y\pm *}_{\ell' m'} \rangle.
    \label{eqn:parity_pair_even}
\end{equation}

If parity is conserved, the expected value of these products must be the same before and after the parity inversion.
This imposes that it has to be zero when $\ell+\ell'$ is odd:
\begin{equation}
    \langle \field^{X+}_{\ell m} \field^{Y+*}_{\ell'm'} \rangle
    = \langle \field^{X-}_{\ell m} \field^{Y-*}_{\ell'm'} \rangle
    = 0, \quad \ell + \ell' \mbox{ odd}.
    \label{eqn:parity_consequence_even}
\end{equation}
Similarly, for two fields having opposite parities,
\begin{equation}
    \langle \field^{X\pm}_{\ell m} \field^{Y\mp *}_{\ell' m'} \rangle \to (-1)^{\ell+\ell'+1} \langle \field^{X\pm}_{\ell m} \field^{Y\mp *}_{\ell' m'} \rangle\,;
    \label{eqn:parity_pair_odd}
\end{equation}
so, if parity is conserved, the expectation values of products of harmonic coefficients must vanish when $\ell+\ell'$ is even:
\begin{equation}
    \langle \field^{X+}_{\ell m} \field^{Y-*}_{\ell'm'} \rangle
    = \langle \field^{X-}_{\ell m} \field^{Y+*}_{\ell'm'} \rangle
    = 0, \quad \ell + \ell' \mbox{ even}.
    \label{eqn:parity_consequence_odd}
\end{equation}

We see that parity invariance alone does not cause correlations between observables of different parity to vanish, it merely forces them into parity-even combinations of harmonic coefficients.
Thus, only combinations with $\ell + \ell'$ odd survive, meaning that the diagonal elements must vanish, but the off-diagonal terms may be non-zero.
It is only in combination with statistical isotropy that parity invariance forces all correlations between observables of opposite parity to vanish, since in this case the product of fields must satisfy both \eqref{eqn:SICdiagonal} and \eqref{eqn:parity_consequence_odd}.

However, statistical anisotropy typically (though not always) goes hand-in-hand with the violation of translation invariance (i.e., with statistical inhomogeneity), which necessarily means that parity is violated at generic locations. We will explore different scenarios in which various topologies break one or more of these symmetries, and their effects on the correlations of CMB polarization anisotropies.

The observed violation of statistical isotropy, therefore, radically changes the expected correlations between harmonic coefficients of observables. If parity is violated, the elements of the correlation matrix $C^{XY}_{\ell m \ell' m'}$ are unconstrained. But even if parity is preserved, only about half of the elements of the correlation matrix $C^{XY}_{\ell m \ell' m'}$ are forced to vanish.

\section{Cosmic topology}\label{sec:topology}

\subsection{Brief introduction to cosmic topology}
The topology and the geometry of the Universe are closely related, but one does not fully determine the other. 
There is a widespread misconception that having an FLRW metric implies that the spatial sections of the manifold have to be isometric to $E^3$ if the metric is flat ($k=0$), to $S^3$ if spherical ($k=+1$), or to $H^3$ if hyperbolic ($k=-1$). 
However, these are not the only manifolds compatible with an FLRW universe \cite{Thurston:1982zz}. 
Even in the flat case, there are $18$ possible topologies that can yield different observable effects \cite{Lachieze-Rey:1995qrb,Luminet:2007xm, COMPACT:2023rkp}, and each of those topologies has associated real parameters whose values affect observables.

Non-trivial topology is a natural way to introduce symmetry violations in the Universe without changing the local physics, as it does not change the metric or the interactions between the fields. Topology only enters the action via changing the integration domain, or equivalently, by setting non-trivial boundary conditions. Interestingly, non-trivial topology can introduce global violations of homogeneity, isotropy, and parity, just by restricting the fields that can exist in these manifolds \cite{COMPACT:2023rkp}. The ultimate reason for these symmetry breakings is that the isometry group of spacetime not only depends on the metric, which is a local notion but also on the manifold where that metric is defined. In the general case, the isometry group of a multi-connected Riemannian manifold (and therefore its possible symmetries) is strictly smaller than the isometry group of its \emph{covering space}.

There are observational constraints on topology from the CMB temperature, mainly from the so-called circles in the sky \cite{Cornish:2003db, ShapiroKey:2006hm, Vaudrevange:2012da, Planck:2013lks, Planck:2015igc}. These constraints amount, approximately, to a lower limit on the shortest closed loop around the Universe through an observer at our location in the Universe. In particular, this loop must be larger than the diameter of the last-scattering surface (LSS) of CMB photons. As we have argued recently \cite{COMPACT:2022nsu,COMPACT:2023rkp,COMPACT:2022gbl, COMPACT:2024dqe}, this constraint still allows for a wide variety of possibilities for topology that can be detected through its influence on the statistics of observables (of the CMB and other probes of cosmic fluctuations).

The effects of topology on the eigenmodes of the spin-0 (scalar) Laplacian, and thence on the CMB temperature fluctuations and its auto-correlations, have been extensively studied (see, e.g., \rcite{Inoue:1998nz, Lehoucq:2002wy, Riazuelo:2003ud, Lachieze-Rey:2005nyz, Weeks:2005ka, COMPACT:2023rkp}).
Less focus has been placed on the effects on \emph{E}-mode polarization fluctuations \cite{Riazuelo:2006tb, Bielewicz:2011jz, Aslanyan:2013lsa, Fabre:2013wia}.
The effects of topology on the eigenmodes of the spin-2 (tensor) Laplacian, and then on CMB \emph{B}-modes, along with the tensor contributions to \emph{T}- and \emph{E}-modes, have remained, to the best of our knowledge, unstudied. 
In future papers, we will explore detailed predictions for the complete set of correlations between \emph{T}, \emph{E}, and \emph{B} modes, generated by both scalar and gravitational tensor (spin-2) modes.
In this work, we focus instead on an important and maybe unexpected consequence of topology in the tensor modes: the production of \EB correlations due to broken symmetries in the simplest Euclidean manifolds without needing microphysical parity violation (even in topologies that preserve global parity).

\subsection{From the covering space to the torus: breaking isotropy}

We explain now how different topologies can affect the observation of parity-odd observables such as \EB correlations. 
We start with the standard infinite Euclidean space, $\Espace$. 
This space is the covering space for all other manifolds with flat geometry but non-trivial topology. 
It is labeled \E{18} in the conventional classification of flat topologies. 
We compare \E{18} with the three first non-trivial Euclidean topologies: \E{1}, the simple 3-torus; \E{2}, the half-turn space; and \E{3}, the quarter-turn space. 
Each of them introduces a new effect on the global symmetries and, in turn, on the observation of \EB correlations.

Plane waves $e^{i\vec{k}\cdot\vec{x}}$, i.e., Fourier modes, are well known to be a complete basis of eigenmodes of the scalar Laplacian in the covering space, \E{18}.
Tensor modes are similarly simple, 
\begin{equation}
    \label{eqn:EuclideanFourierBasis}
    \Upsilon^{\E{18}}_{ij,\vec{k}}(\vec{x},\lambda) = e_{ij}(\unitvec{k},\lambda) e^{i\vec{k}\cdot\vec{x}},
\end{equation}
where $e_{ij}$ is a component of a polarization tensor, which is symmetric, traceless, and transverse (i.e., orthogonal to $\unitvec{k}$). 
There are two independent polarizations, which we will label $\lambda=\pm2$, for the two possible helicities.

The adiabatic tensor perturbation field around the FLRW background at time $t$ can be fully characterized \cite{Tsagas:2007yx, Brechet:2009fa} by the symmetric, traceless, and transverse tensor $\mathcal{D}_{ij}(\vec{x}, t)$, which can be represented in terms of the helicity $\pm 2$ eigenmodes \cite{Weinberg:2008}.
For \E{18},
\begin{equation}
    \mathcal{D}_{ij}(\vec{x},t)=\sum_{\lambda=\pm 2} \int \frac{\dderiv^3 k}{(2\pi)^3}  \mathcal{D}(\vec{k},\lambda,t) \Upsilon^{\E{18}}_{ij,\vec{k}} (\vec{x}, \lambda) .
\end{equation}
Here $\mathcal{D}(\vec{k}, \lambda, t)$ is the time-dependent amplitude for each wavenumber $\vec{k}$ and helicity $\lambda$ corresponding to the eigenmode $\Upsilon^{\E{18}}_{ij, \vec{k}} (\vec{x}, \lambda)$ in the covering space.

Translation invariance of the covering space ensures the statistical independence of modes of different $\vec{k}$ (except, of course, $\vec{k}$ and $-\vec{k}$, which are conjugate because of the reality of the field).
The covariance of $\mathcal{D}(\vec{k},\lambda,t)$ coefficients can be calculated at any time $t$.
In particular, the primordial covariance ($t=0$) is 
\begin{equation}
    \label{eqn:tensor2pcf}
  \langle \mathcal{D}(\vec{k},\lambda, 0) \: \mathcal{D}^*(\vec{k}',\lambda', 0)\rangle = \delta_{\lambda \lambda'} \frac{\pi^2 \mathcal{P}^{T}(k)}{2 k^3}  (2\pi)^3 \delta^{(3)}(\vec{k}-\vec{k}')\,. 
\end{equation}
Here, $\mathcal{P}^{T}(k)$ represents the primordial power spectrum of tensor modes.
The Dirac delta function $\delta^{(3)}(\vec{k}-\vec{k}')$ arises from the translation invariance of the isometry group of the $E^3$ geometry: different Fourier modes are uncorrelated.
The factor $\delta_{\lambda \lambda'}$ is a consequence of the orientability of \E{18}.
The dependence of $\mathcal{P}^{T}(k)$ solely on the magnitude of $\vec{k}$ stems from rotational invariance of the differential operator.
This implies that the power spectrum is a function only of the eigenvalues of the differential operator in the Lagrangian—specifically, the spin-2 Laplacian eigenvalue $-k^2$.
Parity invariance further necessitates that the expression is independent of $\lambda$.
It is assumed, of course, that the isometries of the geometry are respected by the microphysical processes that generated the fluctuations.

We move now to the simple 3-torus, \E{1}. It can be considered a tiling of \E{18} by parallelepipeds, with opposite faces identified. This identification is equivalent to periodic boundary conditions, which have the effect of discretizing the allowed wavevectors $\vec{k}$. 
In particular, if $\vec{T}_1$, $\vec{T}_2$, and $\vec{T}_3$ are the three linearly independent translations that define the identification (i.e., the parallelepiped), the allowed wavevectors must satisfy
\begin{equation}
\label{eqn:allowed}
    \vec{k}_{\vec{n}}\cdot\vec{T}_i = 2\pi n_i\,,\quad i=1,2,3,
\end{equation}
for any triplet of integers $\vec{n}=(n_1,n_2,n_3)$. The tensor modes are now 
$\Upsilon^{\E{1}}_{ij,\vec{n}}(\vec{x},\lambda)$, where the continuous \E{18} label $\vec{k}$ is replaced by the discrete \E{1} label $\vec{n}$.

Tensor modes (like scalar and vector modes) are no longer isotropic in \E{1} because the only allowed modes are the ones with wavevectors $\vec{k}_{\vec{n}}$ on the lattice, which is not invariant under arbitrary rotations.
The helicity $\lambda$ remains a good label for \E{1} eigenmodes because \E{1} is an orientable manifold.
The \E{1} lattice preserves parity, which takes $\vec{k}$ to $\vec{-k}$.
The amplitudes of the two modes $\Upsilon^{\E{1}}_{ij,\vec{n}}(\vec{x},\lambda)$ and $\Upsilon^{\E{1}}_{ij,-\vec{n}}(\vec{x},\lambda)$  remain perfectly correlated by the reality of the tensor field.

The analog of \cref{eqn:tensor2pcf} for  \E{1} is

\begin{equation}
    \label{eqn:tensor2pcfdiscrete}
  \langle \mathcal{D}^{\E{1}}(\vec{k_\vec{n}},\lambda, 0)  \: \mathcal{D}^{\E{1}*}(\vec{k}_{\vecnp},\lambda',0)\rangle
  = \delta_{\lambda \lambda'} \frac{\pi^2 \mathcal{P}^{T}(k_{\vec{n}})}{2 k_{\vec{n}}^3 } V_{\E{1}} \delta_{\vec{n}\vecnp} \,. 
\end{equation}
When converting from the infinite-volume covering space \eqref{eqn:tensor2pcf} to the finite-volume  compact space \E{1}, the expression \( (2\pi)^3 \delta^{(3)}(\vec{k}-\vec{k}') \delta_{\lambda\lambda'}\) is replaced by \( V_{\E{1}} \delta_{\vec{n}\vec{n}'}\delta_{\lambda\lambda'} \), i.e., 
\( \vec{n} \) and $\lambda$ label the \emph{independent} eigenmodes.
\( \mathcal{P}^{T}(k_{\vec{n}}) \) remains a function only of the Laplacian eigenvalues, $-k_{\vec{n}}^2$, or equivalently of
the magnitude of the wavevectors $\vec{k}_{\vec{n}}$.
We retained from the covering space the result that the amplitudes of eigenmodes with different Laplacian eigenvalues are independent Gaussian random variables.\footnote{
     If the $T_i$ encode an accidental symmetry, for example, they are of equal length and orthogonal so that the fundamental domain is a cube, then one can have accidental degeneracies of eigenmodes with different $\vec{n}$, which would allow different choices of ``representative'' eigenmodes.  
     This is exactly as in the covering space, where the rotational invariance implies that if the amplitudes of Fourier modes are independent Gaussian random variables then so are the amplitudes of spherical Bessel functions times spherical harmonics. 
     We lose no generality by ignoring this possibility of accidental degeneracy in the case of \E{1} and the other compact flat manifolds.
}
This again is a consequence of translation invariance for \E{1}, which, like \E{18}, is homogeneous.

Translating \cref{eqn:tensor2pcfdiscrete} into microphysics terms, topology affects only the fields' boundary conditions and leaves the differential operators in the Lagrangian unchanged, assuming that the initial conditions' \emph{probability distribution} maintains the same isotropic and parity-preserving symmetries as local microphysics.
The term \( \mathcal{P}^{T}(k_{\vec{n}}) \), originating from initial conditions and related to microphysics, remains unchanged except for its argument, which is the magnitude of allowed modes in each topology.

The covariance of the eigenmode coefficients given in \cref{eqn:tensor2pcfdiscrete} also applies to the remaining orientable flat topologies \E{2}--\E{6}, with the subtle difference that the eigenmodes are no longer all individual plane waves; rather they are generically linear superpositions of two or more \E{1} eigenmodes with the same helicity $\lambda$ and the same magnitude $k_{\vec{n}}$ of their wavevectors, and their directions related by the generators of the topology.
$\vec{k}_{\vec{n}}$ is the wavevector of one of those plane waves chosen to label the eigenmode.
In non-orientable flat topologies, the expression is analogous but we expect that the eigenmodes mix plane waves with different $\lambda$.
Unlike \E{18} and \E{1}, in \E{2}--\E{6} translation invariance can no longer be invoked to prevent correlations between the amplitudes of Laplacian eigenmodes with different eigenvalues.
Yet, this remains a standard assumption of the field, which we adopt even as we intend to interrogate it further in future work.

The implications of breaking symmetries for \EB correlations in the CMB polarization can now be understood by considering the first three compact Euclidean topologies.
Starting with \E{1}, even without breaking parity conservation, the violation of statistical isotropy in an \E{1} manifold means that the auto-correlation of harmonic coefficients of \emph{E} or \emph{B} will not be diagonal in $\ell$ or $m$.
As seen from \eqref{eqn:parity_consequence_even}, for $XY=EE$ or \emph{BB},
\begin{equation}
    C^{XY, \E{1}}_{\ell m \ell' m'}
            \begin{cases}
                \neq 0, &\ell+\ell' \mbox{ even,} \\ 
                 = 0, &\ell+\ell' \mbox{ odd}.
            \end{cases}
\end{equation}
Further, the cross-correlation of harmonic coefficients of \emph{E} and \emph{B} will not be zero. Instead, as seen from \eqref{eqn:parity_consequence_odd}, we have
\begin{equation}
    C^{EB, \E{1}}_{\ell m \ell' m'} 
            \begin{cases}
               \neq 0, &\ell+\ell' \mbox{ odd,} \\ 
                = 0, &\ell+\ell' \mbox{ even}.
            \end{cases}
\end{equation}
In the next section, we shall see examples of these matrices and how they satisfy the relations expected from the symmetries of the problem. 

\section{Results}\label{sec:results}

In this section, we compute several examples of the covariance matrix $C^{XY, \E{i}}_{\ell m \ell' m'}$ for $XY \in \{EE, BB, EB\}$, produced by tensor (spin-2) perturbations in the Euclidean topologies \E{1}--\E{3}.
The complete equations used in these computations will be detailed in an upcoming work \cite{COMPACT3b}, where we will provide comprehensive calculations for all orientable Euclidean topologies.
The structures of these matrices clearly show the effects of breaking the symmetries discussed above.
Through use of the Kullback-Leibler (KL) divergence \cite{kullback1951, kullback1959information}, we show that information is contained in these structures.

\subsection{\E{1}, \E{2}, and \E{3}: different degrees of symmetry breaking}

Though all Euclidean topologies are affected, it is illustrative to restrict our attention to three compact orientable Euclidean topologies to show the progressive breaking of the symmetries of \E{18}.
To compute these covariance matrices, we use the power spectrum (and transfer functions) of the best-fit {\it Planck} cosmology \cite{Planck:2018vyg} with a tensor spectral index of \( n_T=-0.0128 \). 
To clarify the effects of topology, and minimize the role of specific parameter assumptions, we consider \emph{rescaled} covariance matrices where each element is given by
\begin{equation}
    \Xi^{\E{i}, XY}_{\ell m\ell'm'} \equiv \frac{C^{\E{i}, XY}_{\ell m\ell'm'}} {\sqrt{C^{\E{18}, XX}_{\ell}C^{\E{18}, YY}_{\ell'}}}\,.
\end{equation}

\begin{figure}
    \centering
    \textbf{Cubic $\E{1}, \ L = 0.8 L_{\mathrm{LSS}}$}\par\medskip
    \includegraphics[width=1\linewidth]{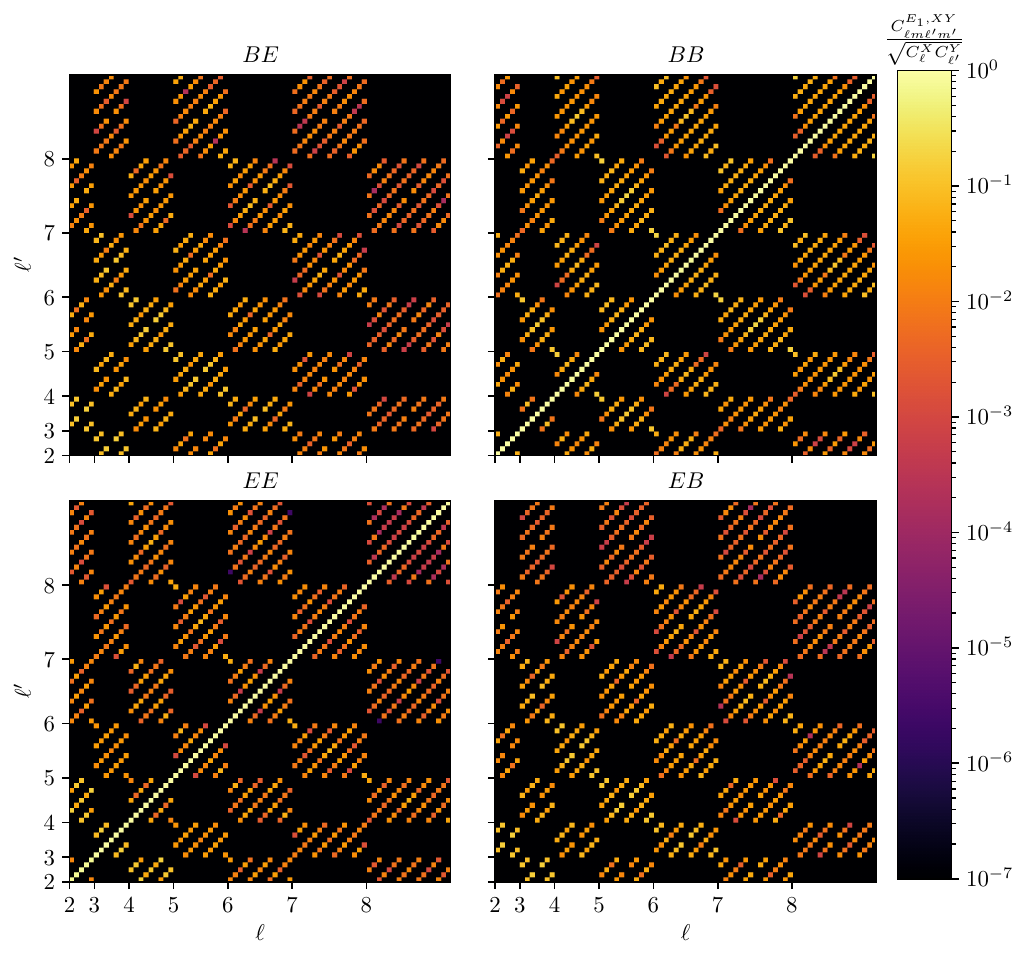}
    \caption{Absolute values of the rescaled CMB covariance matrices for \emph{EE} (bottom left), \emph{BB} (top right), and the cross-covariances \emph{BE} and \emph{EB} (top left and bottom right). They are computed at low multipoles $\ell\leq 8$, and using ``$\ell$ ordering'' for the cubic, untilted \E{1}  with $L = 0.8 L_{\mathrm{LSS}}$ and for an on-axis observer. 
    Here, $L$ is the length of the translations generating the topology.}
    \label{fig:cov_matrix_E1}
\end{figure}

\begin{figure}
    \centering
    \textbf{Cubic $\E{3}, \ L = 0.8 L_{\mathrm{LSS}}$}\par\medskip
    \includegraphics[width=1\linewidth]{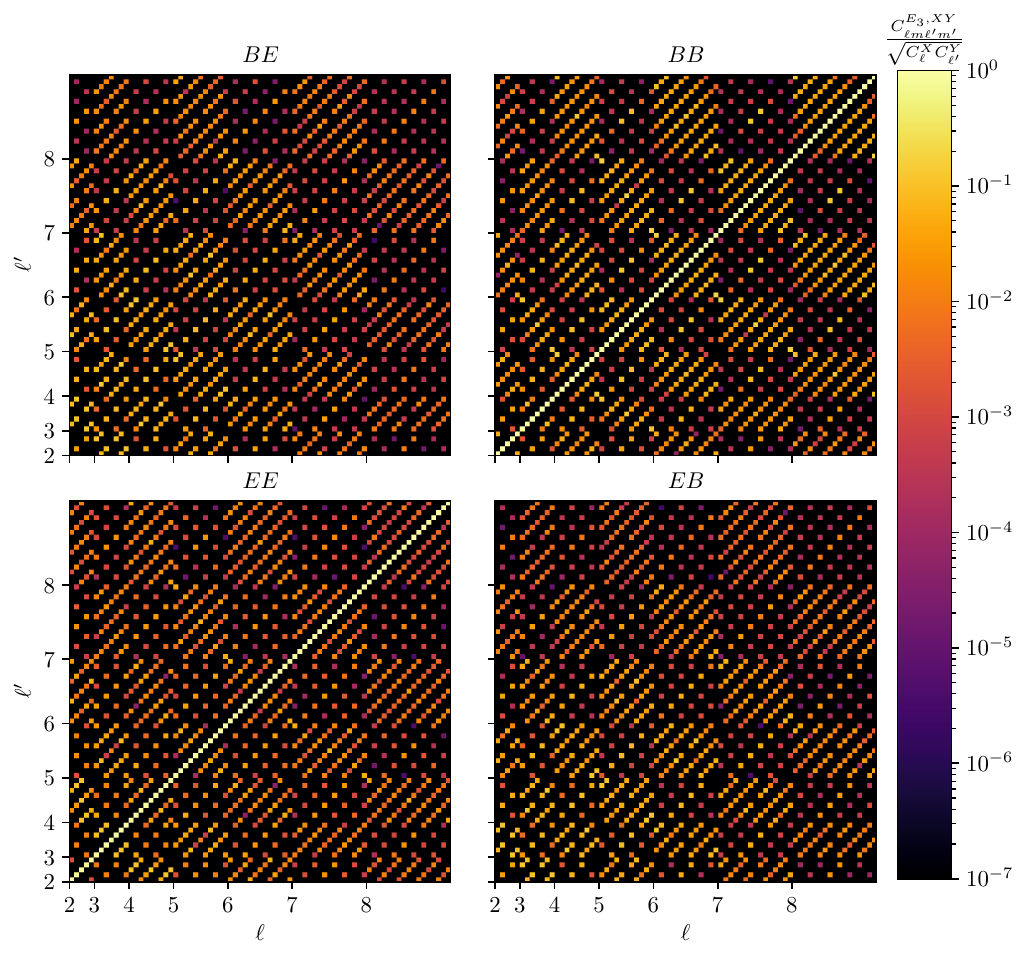}
    \caption{ 
        As in \cref{fig:cov_matrix_E1}, but for the cubic, untilted \E{3}  with $L = 0.8 L_{\mathrm{LSS}}$ and an on-axis observer.
    }
    \label{fig:cov_matrix_E3}
\end{figure}

\begin{figure}
    \centering
    \textbf{Cubic $\E{2}, \ L = 0.8 L_{\mathrm{LSS}}$}\par\medskip
    \includegraphics[width=1\linewidth]{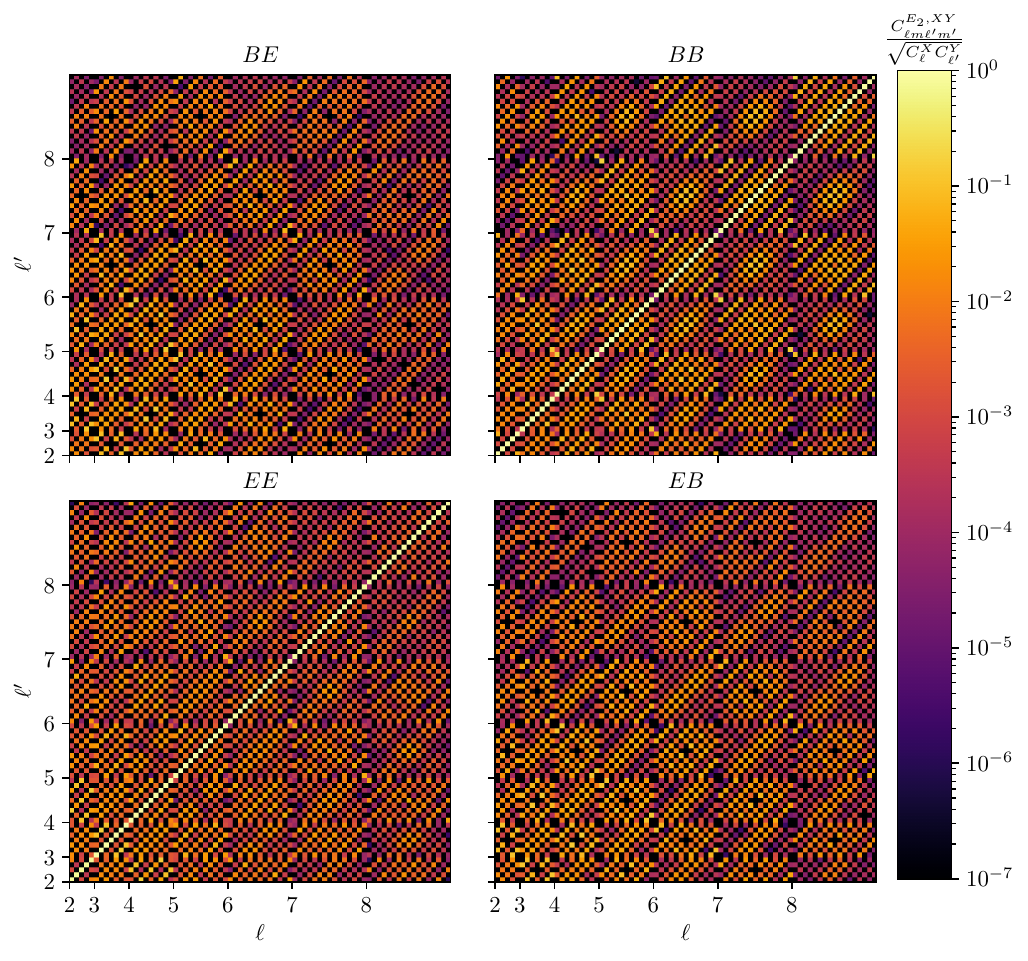}
    \caption{
        As in \cref{fig:cov_matrix_E1}, but for the cubic, untilted \E{2} with $L = 0.8 L_{\mathrm{LSS}}$ and an off-axis observer at $\vec{x}_0 = (0.36 \ L_\mathrm{LSS},\ 0.36 \ L_\mathrm{LSS},\ 0)$. 
    }
    \label{fig:cov_matrix_E2}
\end{figure}

\cref{fig:cov_matrix_E1,fig:cov_matrix_E3,fig:cov_matrix_E2} show the modulus of $\Xi^{\E{i}, XY}_{\ell m\ell'm'}$ (as it is a complex matrix).
We present it in ``$\ell$ ordering'', i.e., in increasing order of the multipole $\ell$, and, within each multipole, in increasing order of $m$ from $m=-\ell$ to $m=+\ell$.
Note that $\Xi^{\E{i}, XY}$ is independent of the choice of tensor-to-scalar ratio \emph{r} because we only include the contributions of tensor fluctuations in the covariance matrices.
Once we incorporate both the scalar and tensor contributions in the same covariance matrices in \rcite{COMPACT3c}, studying the dependence of the correlation signal on $r$ and $n_T$ will be important.

In \cref{fig:cov_matrix_E1}, we display the \emph{EE}, \emph{BB}, and \EB covariance matrices for a cubic \E{1}, where the length of the translation vectors is $L=0.8 L_{\mathrm{LSS}}$, with $L_{\mathrm{LSS}}$ the diameter of the last-scattering surface.
It can be seen that $(\ell+\ell')$-odd correlations in \emph{EE} and \emph{BB} vanish, whereas $(\ell+\ell')$-even correlations in \EB vanish.
This is the expected behavior in a space that breaks statistical isotropy but preserves parity as represented in \eqref{eqn:parity_consequence_even} and \eqref{eqn:parity_consequence_odd}.
We note that the correlations for modes with $(m-m')\mod{4}\neq 0 $ also vanish. However, this is a consequence of the accidental cubic symmetry and our choice to orient our axes parallel to the edges of the cube.
This additional vanishing does not occur in general.

The half-turn space, \E{2}, can be understood similarly to \E{1}: we consider a similar cubic fundamental domain and identify its opposite sides. However, unlike for \E{1}, one pair of faces is identified with a half turn ($\pi$ rotation).
Conventionally, this pair of faces is chosen to lie in the $xy$-plane. 
\E{2} is no longer homogeneous: the axis of rotation is a preferred location. 
For an observer located on the axis of the rotation, parity is conserved, and the correlations have the same symmetries as those of \E{1}.
When the observer is off the axis of rotation, parity is violated in \E{2}: the direction toward the axis of rotation is distinct from the direction away from the axis. 

The \E{3} topology is similar to \E{2}, but now with one pair of opposite faces identified with a quarter turn ($\pi/2$ rotation), instead of a half-turn rotation.
Again, the convention is to choose the identification to be a translation along the $z$-direction accompanied by a rotation by $\pi/2$ about the $z$-axis.
Since a rotation by $\pi/2$ is not equivalent to a rotation by $-\pi/2$, the space has a handedness and global parity is violated everywhere, even for an on-axis observer.

The normalized covariance matrix for an on-axis observer in \E{3} is shown in \cref{fig:cov_matrix_E3}.
There are now non-zero correlations for all of \emph{EE}, \emph{BB}, and \EB, regardless of whether $\ell+\ell'$ is even or odd. We note that the \EB correlations for this observer have non-zero elements in the diagonal.
There are many null correlations in the correlation matrices, as the observer being on-axis imposes extra symmetries. 

We focus next on off-axis observers to show how the violation of symmetries is reflected in the correlation matrices.
In \cref{fig:cov_matrix_E2}, we show these correlation matrices for an off-axis observer in \E{2}.
The effect of the global parity violation in \E{2} is as expected: $(\ell+\ell')$-odd correlations no longer vanish for \emph{EE} and \emph{BB}, and $(\ell+\ell')$-even correlations no longer vanish for \EB.
All $(\ell,\ell')$ blocks have non-zero elements. 
However, notice that the particular properties of \E{2} still prevent \EB correlations that are diagonal in $(\ell,m)$. For an off-axis observer in \E{3}, all possible \EB correlations are present (except that $\lambda$ and $\lambda'$ still do not mix).

We have shown that flat topologies can produce diagonal \EB correlations, as well as non-diagonal ones. Given that these correlations cannot be produced by scalar modes, their detection would open a new window to study tensor perturbations and the tensor-to-scalar ratio $r$, in a way that is not possible in the covering space.
We have shown the results of the first three topologies as they nicely illustrate the progression in the symmetry violation caused by topology; the remaining compact topologies break all the global symmetries in a way similar to \E{3}.

Finally, we emphasize that the full covariance matrix should be considered to extract information about the topology, not only the diagonal part, especially for the cross-correlation \EB. 
Looking only at the diagonal part of the covariance matrix could make this signal indistinguishable from parity-violating microphysical effects, such as cosmic birefringence. The effects of cosmic birefringence have been extensively studied, especially in the diagonal part of the covariance matrix \cite{Kitajima2022,nakatsuka2022,namikawa2024,Zagatti:2024jxm}. In a follow-up paper, we will compare the predictions of birefringence and non-trivial topologies for \EB correlations, highlighting their distinct features and exploring whether their signals could be degenerate in specific scenarios \cite{COMPACT3c}.

\subsection{Information content and KL divergence}

We have shown in the previous section that the covariance matrix of the polarization $a_{\ell m}$ is qualitatively different in different topologies. However, this covariance matrix is not directly observable.
A key question is whether these different covariance matrices will produce distinguishable observables, i.e., whether the probability distribution of the observables is different enough.
In order to study this question, we can make use of the KL divergence. The KL divergence between two probability distributions $p$ and $q$, also known as their \textit{relative entropy}, is defined as
\begin{equation}
    \label{eqn: KL_Divergence}
     D_{KL}(p||q) = \int \mathrm{d}\vec{x}\,p(\vec{x}) \ln\!\left(\frac{p(\vec{x})}{q(\bm{x})}\right)\,.
\end{equation}
In the Bayesian framework, the KL divergence between two distributions is nothing but the expected Bayes factor (i.e., log-likelihood ratio) associated with those distributions assuming the data follows the first distribution. 

\begin{figure}[h]
    \centering
        \centering
        \includegraphics[width=1\textwidth]{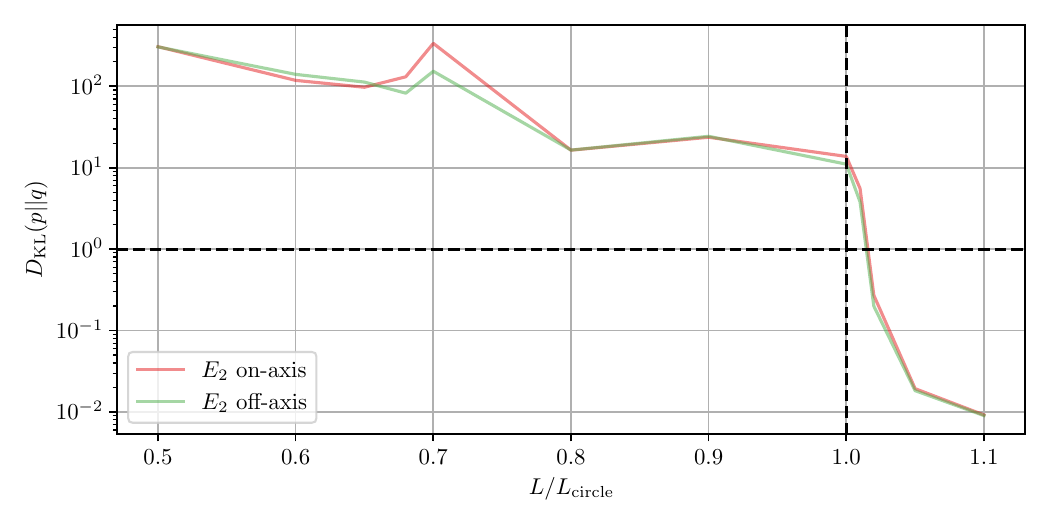}
    \caption{The KL divergence between the flat covering space \E{18} and the Euclidean topology \E{2} for tensor-induced \emph{EE}, \emph{BB}, and \emph{EB} correlations as a function of topology scale. 
    In this analysis, we compare two scenarios for cubic \E{2} where the observer is either on-axis or off-axis at \(\vec{x}_0 = (0.45L, 0.45L, 0.0)\), with \(L\) representing the size of the cubic box. The $x$-axis shows \(L/L_{\mathrm{circle}}\), where \(L_{\mathrm{circle}}\) is the smallest value of $L$ for which there is no closed-loop geodesic through the observer of length less than $L_{\mathrm{LSS}}$
    \cite{COMPACT:2023rkp,COMPACT:2022nsu}. 
    For cubic \E{1}--\E{3}, \(L_{\mathrm{circle}} = L_{\mathrm{LSS}}\) regardless of observer position.
    We see that for $L/L_{\mathrm{circle}}\lesssim1$ there is significant information in the \emph{E} and \emph{B} correlations.
    }
    \label{fig:KL main}
\end{figure}

Computing the KL divergence between the different probability distributions that CMB temperature and polarization fluctuations would follow in different topologies is a common procedure to compare models in cosmic topology \cite{Kunz:2007kw,Fabre:2013wia,Planck:2015gmu,COMPACT:2022gbl,COMPACT:2023rkp}. Given that the $a_{\ell m}^X$ coefficients of the CMB follow a zero-mean Gaussian distribution, we can simplify \cref{eqn: KL_Divergence} to
\begin{equation}
D_{KL}(p||q) = \frac{1}{2}\sum_i(\ln|\lambda_i|+\lambda_i^{-1}-1)\,,
\end{equation}
where the $\{\lambda_i\}$ are the eigenvalues of the matrix 
\begin{equation*}
 C_{\ell m\ell' m'}^{XY,\,p} \, (C_{\ell m\ell' m'}^{XY,\, q})^{-1}\,.
\end{equation*}
 The distribution $q$ is typically taken to be that of the observable in the covering space (which has the advantage of having a diagonal covariance matrix) whereas $p$ is usually the distribution of the observable in the non-trivial topology under study. A value of $D_{KL}(p||q)\geq 1$ is typically considered as the threshold for both topologies to be distinguishable, assuming data actually follows distribution $p$. This is equivalent to saying that the expected value of the (natural) logarithm of the Bayes ratio is $1$, typically considered weak evidence in favor, according to the widely used Jeffrey's scale.
This value provides a quantitative measure of the detectability of non-trivial topology in an ideal experiment with no noise, foreground emission, or masking. 

In \cref{fig:KL main}, we present the KL divergence between the flat covering space \E{18} and the half-turn space \E{2} for tensor-induced \emph{EE}, \emph{BB}, and \emph{EB} correlations as a function of topology size.
We compare cubic \E{2} with an on-axis observer to cubic \E{2} with an off-axis observer at \(\vec{x}_0 = (0.45L, 0.45L, 0.0)\), where \(L\) is the size of the cubic box. 
The KL divergence for cubic \E{1} and \E{3} with an on-axis observer exhibits similar behavior to that of cubic \E{2} with an on-axis observer.
The detailed behavior of the KL divergence will be explored in \rcite{COMPACT3b}.
The $x$-axis is shown as $L/L_{\mathrm{circle}}$, where $L_{\mathrm{circle}}$ is defined as the smallest size for which no pairs of circles on the CMB sky have matching patterns of temperature fluctuations \cite{Cornish:1996kv, Cornish:1997ab}.
Equivalently, $L_{\mathrm{circle}}$ is the smallest value of $L$ for which there is no closed-loop geodesic through the observer of length less than $L_{\mathrm{LSS}}$. 
In cubic \E{1}--\E{3}, whether the observer is on-axis or off-axis,  $L_{\mathrm{circle}}=L_{\mathrm{LSS}}$.  

A more complete study will appear in \rcite{COMPACT3c}, where we will compute the KL divergence for the full $T$, $E$, and $B$ correlations across different topologies, including both scalar and tensor contributions.
This measure will better reflect the detectability of non-trivial topologies, as it is observationally impossible to disentangle the scalar and tensor contributions.

\section{Conclusions}

We have shown that non-trivial cosmic topology can induce non-zero \EB correlations, as well as additional off-diagonal elements in \emph{EE}, \emph{EB}, and \emph{BB} correlations. Statistical isotropy typically restricts correlations to diagonal terms in harmonic space. By considering multi-connected manifolds, we can break statistical isotropy, homogeneity, and parity globally, without altering the microphysics or introducing new terms into the early or late Universe's Lagrangian. Compared to the covering space, this violation of statistical isotropy allows half of the non-diagonal terms to emerge; the violation of parity for generic observers in generic manifolds eliminates the symmetry protection for the remaining terms.

The detection of non-zero parity-odd correlations in CMB polarization data, especially \EB correlations, emerges as a potent indicator of the Universe's underlying topology.
This evidence, pointing towards a violation of statistical isotropy, underscores the delicate nature of parity in a Universe that may exhibit anisotropic properties.
We have shown that the tensor perturbations can produce observationally different imprints in the CMB polarization for different topologies, at least in an ideal experiment without noise, foregrounds, or masks.
Conversely, our results show that tensor-mode perturbations could more easily be detected in a manifold with non-trivial topology, as it produces both \emph{BB} and \EB correlations, which would help measure the tensor-to-scalar ratio $r$.

Our analysis highlights the crucial role of considering non-trivial cosmic topologies in interpreting CMB data, stressing the need to go beyond the isotropic power spectrum coefficients $C_\ell$. 
In upcoming papers, we will explore the full correlations induced by both tensor and scalar perturbations in both orientable and non-orientable Euclidean manifolds, as well as other topologically non-trivial manifolds with positive and negative curvature.

\acknowledgments
This work made use of the High-Performance Computing Resource in the Core Facility for Advanced Research Computing at Case Western Reserve University.
T.S.P. acknowledges financial support from the Brazilian National
Council for Scientific and Technological Development (CNPq) under grants 312869/2021-5
and 88881.709790/2022-0. Y.A. acknowledges support by the Spanish Research Agency (Agencia Estatal de Investigaci\'on)'s grant RYC2020-030193-I/AEI/10.13039/501100011033, by the European Social Fund (Fondo Social Europeo) through the  Ram\'{o}n y Cajal program within the State Plan for Scientific and Technical Research and Innovation (Plan Estatal de Investigaci\'on Cient\'ifica y T\'ecnica y de Innovaci\'on) 2017-2020, and by the Spanish Research Agency through the grant IFT Centro de Excelencia Severo Ochoa No CEX2020-001007-S funded by MCIN/AEI/10.13039/501100011033.  J.R.E. acknowledges support from the European Research Council under the Horizon 2020 Research and Innovation Programme (Grant Agreement No.~819478).
C.J.C., G.D.S., D.P.M., and A.K.\ acknowledge partial support from NASA ATP grant RES240737; G.D.S. and A.S.\ from DOE grant DESC0009946; G.D.S. and Y.A.\ from the Simons Foundation; G.D.S., Y.A., and A.H.J.\ from the Royal Society (UK); and A.H.J.\ from STFC in the UK\@.
A.T.\ is supported by the Richard S.\ Morrison Fellowship.
G.D.S. and Y.A. thank the INFN (Sezione di Padova), and D.P.M.,  G.D.S., S.A., J.R.E., and A.T. thank the IFT for hospitality where part of this work was accomplished.

\bibliographystyle{utphys}
\bibliography{topology,additional,additional2}

\end{document}